\documentclass[runningheads]{llncs}
\usepackage{graphicx}
\usepackage{comment}
\usepackage{amsmath,amssymb} 
\usepackage{color}
\usepackage{tikz}
\usepackage{booktabs}

\usepackage{times}
\usepackage{amsmath}
\usepackage{amssymb}

\usepackage{enumitem}
\usepackage{url}
\usepackage{authblk}
\usepackage{makecell}
\usepackage{bm}

\usepackage[font=small,skip=0pt]{caption}
\newcommand{\etal}{\textit{et al.}}
\DeclareMathOperator*{\E}{\mathbb{E}}

\begin{document}
\pagestyle{headings}
\mainmatter

\title{An End-to-End Joint Learning Scheme of Image Compression and Quality Enhancement with Improved Entropy Minimization} 


\titlerunning{ }
%
\author{Jooyoung Lee\inst{1,3} \and
Seunghyun Cho\inst{2} \and
Munchurl Kim\thanks{Corresponding author.}\inst{3}}
\authorrunning{ }
%
\institute{Electronics and Telecommunications Research Institute, Korea \\
\email{leejy1003@etri.re.kr} \and
Kyungnam University, Korea \\
\email{scho@kyungnam.ac.kr} \and
Korea Advanced Institute of Science and Technology, Korea\\
\email{mkimee@kaist.ac.kr}}

\maketitle

\begin{abstract}
Recently, learned image compression methods have been actively  studied. Among them, entropy-minimization based approaches have achieved superior results compared to conventional image codecs such as BPG and JPEG2000. However, the quality enhancement and rate-minimization are conflictively coupled in the process of image compression. That is, maintaining high image quality entails less compression and vice versa. However, by jointly training separate quality enhancement in conjunction with image compression, the coding efficiency can be improved. In this paper, we propose a novel joint learning scheme of image compression and quality enhancement, called JointIQ-Net, as well as entropy model improvement, thus achieving significantly improved coding efficiency against the previous methods. Our proposed JointIQ-Net combines an image compression sub-network and a quality enhancement sub-network in a cascade, both of which are end-to-end trained in a combined manner within the JointIQ-Net. Also the JointIQ-Net benefits from improved entropy-minimization that newly adopts a Gussian Mixture Model (GMM) and further exploits global context to estimate the probabilities of latent representations. In order to show the effectiveness of our proposed JointIQ-Net, extensive experiments have been performed, and showed that the JointIQ-Net achieves a remarkable performance improvement in coding efficiency in terms of both PSNR and MS-SSIM, compared to the previous learned image compression methods and the conventional codecs such as VVC Intra (VTM 7.1), BPG, and JPEG2000. To the best of our knowledge, this is the first end-to-end optimized image compression method that outperforms VTM 7.1 (Intra), the latest reference software of the VVC standard, in terms of the PSNR and MS-SSIM.

\end{abstract}
\section{Introduction}
\label{sec:introduction}
\begin{figure}[t]
\begin{center}
\includegraphics[width=1.0\linewidth]{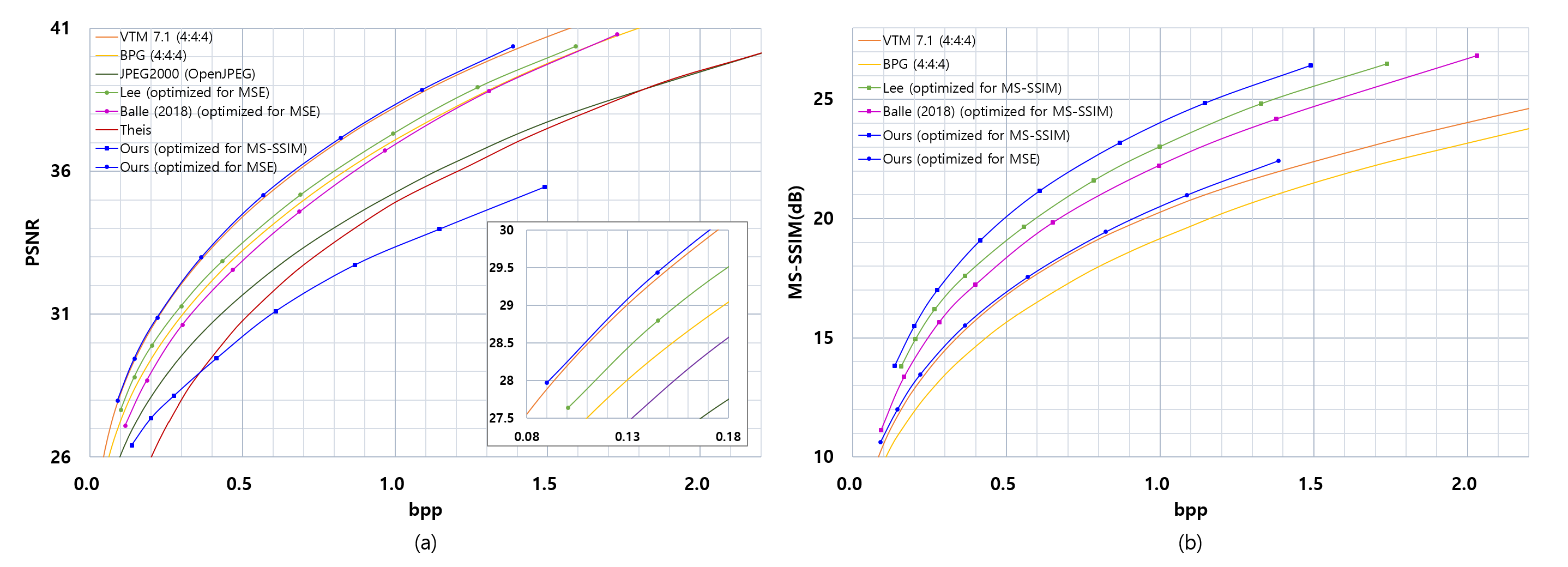}
\end{center}
   \caption{Rate-distortion curves of the proposed method and competitive methods for the Kodak PhotoCD image dataset \cite{KODAK}. The left and right plots represent RD-curves in terms of (a) PSNR and (b) MS-SSIM, respectively. Note that the measured MS-SSIM values are presented in the unit of decibels as in the previous works~\cite{Balle18,Minnen2018,Lee2019} to better distinguish the performance differences.}
\label{fig:experimental_results}
\end{figure}
\begin{figure*}[t]
\includegraphics[width=1.0\linewidth]{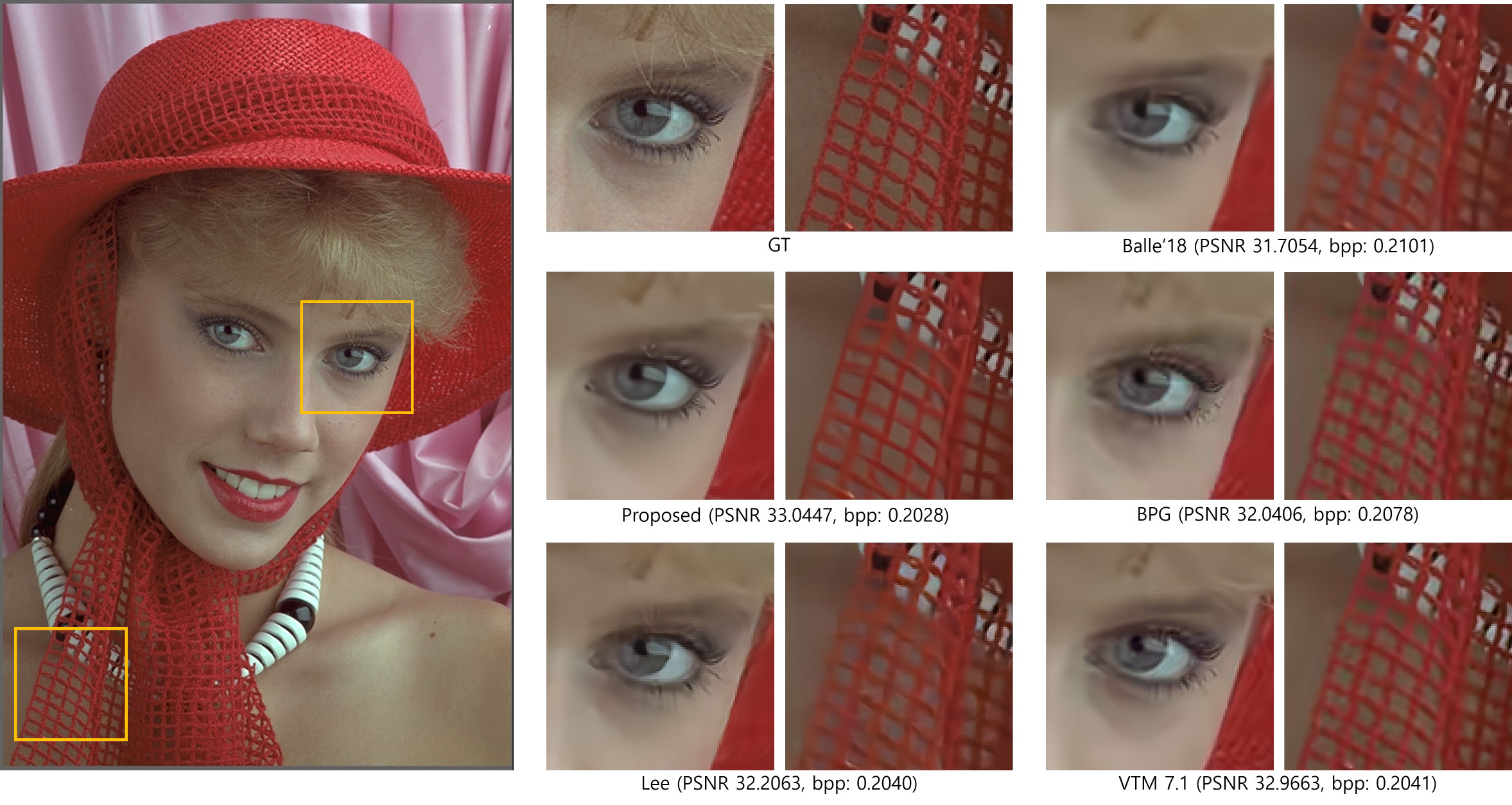}
\caption{Comparison of sample reconstruction images including the ground truth, our method, Lee \etal~\cite{Lee2019}'s approach, Ball{\'{e}} \etal(2018)~\cite{Balle18}'s approach, BPG \cite{BPG}, and VVC Intra (VTM 7.1) \cite{VTM}.}
\label{fig:sample_recon_images}
\end{figure*}

Recently, significant progress in artificial neural networks has led to many groundbreaking achievements in various research fields. In image and video compression domain, a number of learning based studies~\cite{Toderici17,Johnston2018,Balle17,Theis17,Balle18,Lee2019,Minnen2018,Park2019,lu2018dvc} have been conducted. Especially, some latest end-to-end optimized image compression approaches \cite{Lee2019,Minnen2018} based on entropy minimization have already shown better compression performance than those of the existing image compression codecs such as BPG \cite{BPG} and JPEG2000~\cite{Taubman2001}, despite a short history of the field. The basic approach to entropy minimization is to train analysis (encoder) / synthesis (decoder) transform networks allowing them to reduce entropy of transformed latent representations, keeping the quality of reconstructed images as close as possible to the originals. Entropy minimization approaches can be viewed from two different aspects: prior probability modeling and context exploitation. Prior probability modeling is a main element of entropy minimization and allows an entropy model to approximate the actual entropy of latent representations, which plays a key role for both training and actual entropy coding/decoding. 
For each transformed representation, an image compression method estimates the parameters of the prior probability model, based on contexts such as previously decoded neighbor representations or some bit-allocated side informations. A better context can be regarded as the information given to the model parameter estimator, which help predict the distributions of latent representations more precisely. 

The latest two entropy minimization approaches \cite{Lee2019,Minnen2018} achieved noticeable compression performance, but their methods focused on building new entropy models with context exploitation in an autoregressive manner, rather than utilizing the up-to-date architectural techniques. Meanwhile, in the field of quality enhancement, a number of studies have been continuously conducted in an architectural perspective and have shown superior results compared to the traditional methods, as described in Section \ref{sec:related work}. However, there has been few work on jointly taking into account both image compression and quality enhancement in a unified architecture, although worthwhile to restore the compressed images with coding artifacts as close as possible to the quality of uncompressed input. Therefore, in this paper, we propose a novel joint learning scheme that incorporates quality enhancement and image compression so as to allow them to collaborate each other for higher coding efficiency. However, we do not propose any specific quality enhancement network to be incorporated into our new image compression network. Instead, an state-of-the-art (SOTA) method is adopted, so that any advanced method can be combined with our image compression network in the proposed unified joint learning architecture.  

In addition, we present a novel image compression network that incorporates an improved entropy minimization method with GMM-based prior probability modeling and global context exploitation. In terms of prior probability modeling, we adopt a more generalized form with a GMM. The GMM was simply mentioned but was not used in Minnen \etal~\cite{Minnen2018}’s approach where a single Gaussian model was used in their formulation and experiments. In our prior probability modeling, the sub-network for estimating the model parameters of GMM is trained in the course of jointly learning image compression and quality enhancement, yielding the improved estimation accuracy for them.

From a contextual perspective, we define a new type of global context for entropy minimization in an autoregressive manner. The autoregressive approaches \cite{Lee2019,Minnen2018} estimate the distribution of a current latent representation using its adjacent known representations, thus leading to improve the compression performance by removing the correlations between the current latent representation and its neighbors. Although their methods effectively remove the spatial and inter-channel correlations among the transformed representations, our global context exploitation further improves the coding efficiency by removing the remaining spatial correlations across a wider area of each input image, which has been motivated by the known wisdom~\cite{glasner2009super,D.Y.LEE_2017} that exploits self-similarity in input images.

Fig.~\ref{fig:experimental_results} shows the coding efficiency curves for our JointIQ-Net, Versatile Video Coding (VVC) Intra (VTM 7.1~\cite{VTM}), BPG~\cite{BPG}, JPEG2000~\cite{Taubman2001} and the deep learning based SOTA methods~\cite{Lee2019,Balle18,Theis17}. As shown in Fig.~\ref{fig:experimental_results}-(a) and -(b), our JointIQ-Net outperforms all the image compression methods in terms of both PSNR and MS-SSIM. It is noted that our JointIQ-Net is the \textit{first} deep image compression scheme that surpasses the VVC Intra coding. Fig.~\ref{fig:sample_recon_images} visually compares some reconstructed images for our JointIQ-Net and the existing methods~\cite{Lee2019,Balle18,BPG,Taubman2001}. For similar compression ratios, it is clearly shown in Fig.~\ref{fig:sample_recon_images} that our JointIQ-Net yields the reconstructed images of higher quality in both quantitative measures (PSNR) and qualitative viewing. The key contributions of our work are as follows:

\begin{itemize}[leftmargin=15pt,itemsep=2pt,topsep=2pt]
\item[$\bullet$] We first propose a novel end-to-end learning scheme, called JointIQ-Net, that can jointly optimize both image compression and quality enhancement;
\item[$\bullet$] To the best of our knowledge, the JointIQ-Net is the \textit{first} deep image compression scheme that outperforms in terms of both PSNR and MS-SSIM the VVC Intra coding (VTM 7.1~\cite{VTM}) which has been almost finalized for standardization by ISO/IEC MPEG and ITU-T VCEG, also yielding significant improvements over BPG, JPEG2000, and the learned SOTA image compression approaches.
\item[$\bullet$] We propose an improved entropy-minimization method that uses a GMM for prior probability modeling, whose parameters are accurately estimated by the improved estimator trained in the joint optimization of image compression and quality enhancement, yielding the improved coding efficiency;
\item[$\bullet$] To further improve the entropy-minimization method, we utilize global context in estimation of the GMM parameters, which captures a wider context information and helps reduce the spatial correlations between a current latent representation and its neighbors in a non-local extent;
\end{itemize}

\section{Related work}
\label{sec:related work}
Artificial-neural-network (ANN) based image compression approaches can be divided into two folds: First, some approaches \cite{Toderici17,Johnston2018} try to achieve a small number (or ratio) of latent representations while maintaining the original information as much as possible in latent spaces. Based on this concept, Toderici \etal~\cite{Toderici17} introduced a novel image compression method using a fixed number of latent binary representations, which improve the image quality in an progressive manner. Then Johnston \etal~\cite{Johnston2018} enhanced the network operation method of Toderici \etal~'s network to achieve better coding efficiency; Second, some other approaches \cite{Balle17,Theis17,Agustsson2017,Balle18,Lee2019,Minnen2018} minimize the entropy of the latent representations, which transforms them to have low entropy to be represented in a small number of bits by using their own entropy models. Ball{\'{e}} \etal(2017)~\cite{Balle17} and Theis \etal~\cite{Theis17} introduced a new image compression method based on entropy minimization. Ball{\'{e}} \etal(2018)~\cite{Balle18} enhanced the entropy model by adopting a hierarchical prior model for estimating standard deviations of the latent representations in an input-adaptive manner, whereas the first two approaches \cite{Balle17,Theis17} train their image compression networks with their prior model parameters fixed during inference. Minnen \etal~\cite{Minnen2018} and Lee \etal~\cite{Lee2019} utilize adjacent regions of known latent representations as additional contexts for the parameter estimation of prior models, based on the idea that entropy-coding and decoding process can be conducted in an autoregressive manner (e.g. a raster scanning order) and spatially adjacent representations tend to have high correlations. Both approaches enhanced the compression performance and obtained better results than BPG \cite{BPG} that is the image compression codec based on HEVC (ISO/IEC 23008-2, ITU-T H.265)~\cite{HEVC}.

Meanwhile, ANN-based image restoration, such as super resolution (SR) and denoising, has become an indispensable method by far surpassing handcrafted algorithms. Kim \etal~\cite{Kim_2016_VDSR}'s approach, first introduced a deep network architecture based on residual learning for SR, named VDSR, and obtained substantial boost in SR performance. Zhang \etal~\cite{Zhang_2018_CVPR}'s approach has achieved further improvement by exploiting residual dense blocks (RDBs), each of which comprises densely connected convolutional layers and a local skip connection. Kim \etal~\cite{Kim_GRDN_2019}'s approach, grouped residual dense network (GRDN), has extended the previous work by grouping multiple RDBs, named grouped residual dense blocks (GRDBs), and arranged the multiple GRDBs in the network. Furthermore, they incorporate a more deeper architecture that allows the convolutional layers to process down-scaled representations, and also adopt spatial and channel-wise attention layers. Based on this enhanced architecture, they has obtained the state-of-the-art performance in the image denoising task. Recently, Cho~\etal~\cite{Cho_2019_CVPR_Workshops}'s approach has utilized GRDN~\cite{Kim_GRDN_2019} for reducing artifacts caused by a new image codec, which is the intra coding of VVC ~\cite{VVC}, under standardization, and they have achieved a noticeable quality improvement. However, GRDN in Cho \etal~\cite{Cho_2019_CVPR_Workshops}'s approach has been separately optimized against the image codec.
\begin{figure}[t]
\begin{center}
\includegraphics[width=0.6\linewidth]{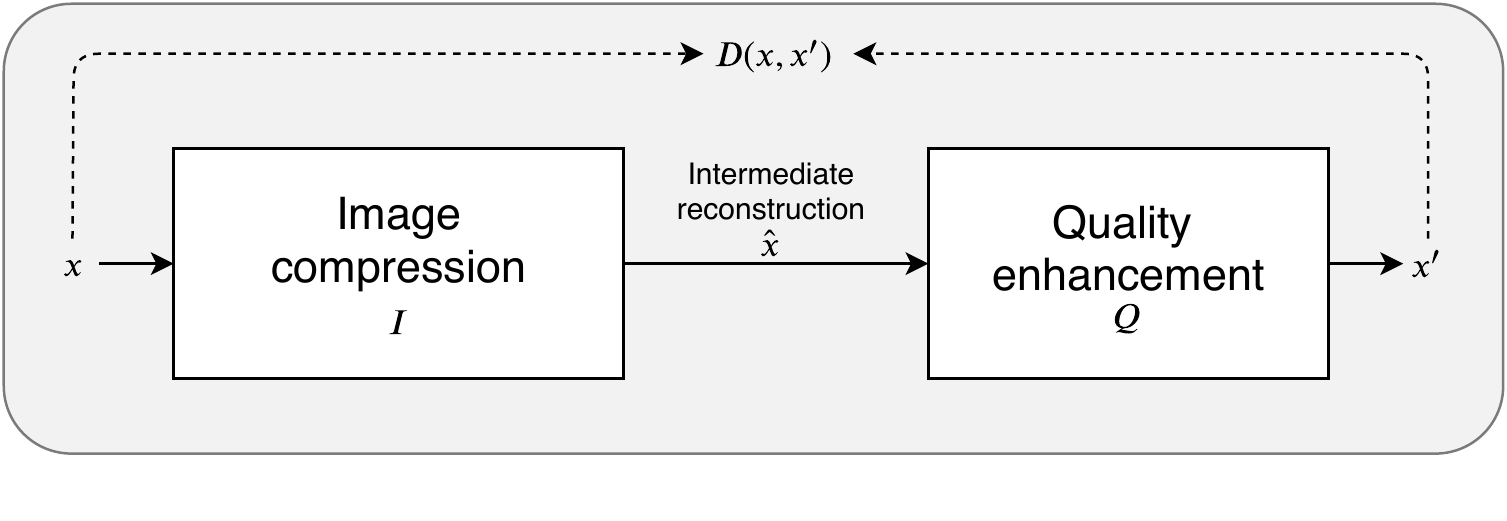}
\end{center}
   \caption{Joint learning scheme of image compression and quality enhancement.}
\label{fig:joint_opt_framework}
\end{figure}

\section{Proposed network architecture}
\subsection{Joint learning scheme of image compression and quality enhancement}
\label{sec:Joint_opt}

\begin{figure}[t]
\begin{center}
\includegraphics[width=1.0\linewidth]{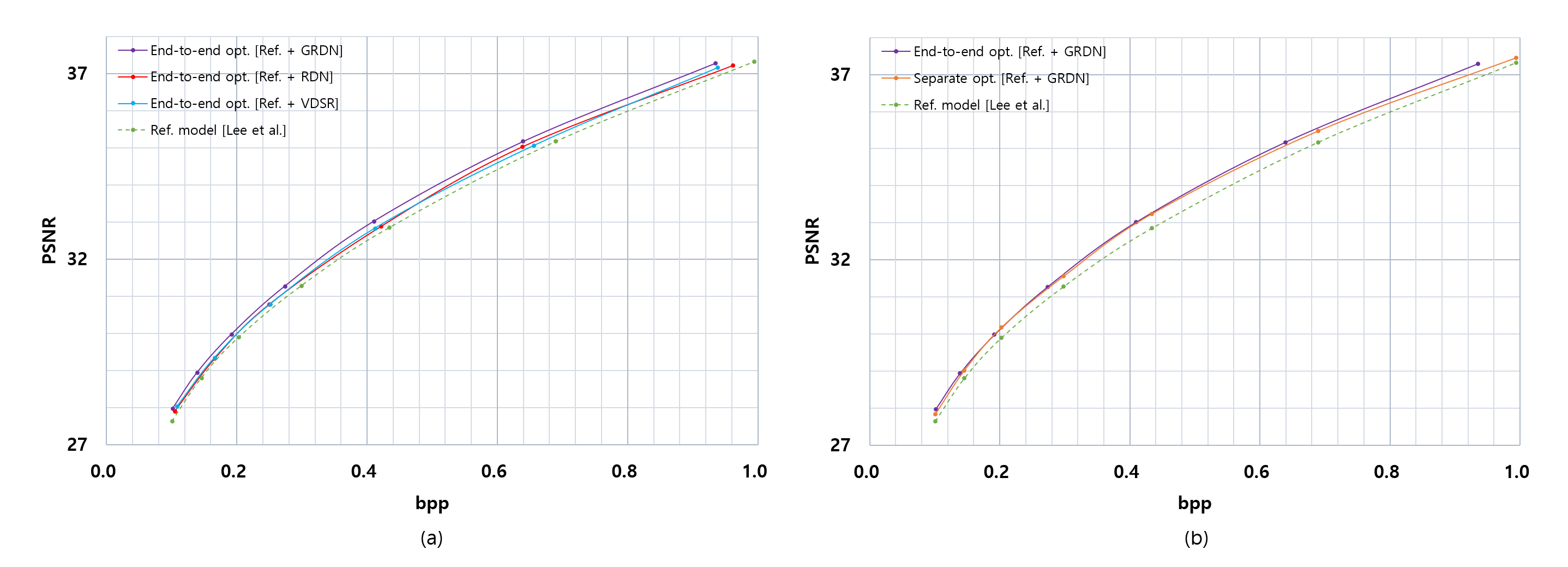}
\end{center}
   \caption{Compression performance for cascades of a reference image compression network and GRDN over PSNR vs. bpp. (a) the reference image compression network~\cite{Lee2019} jointly optimized with various quality enhancement methods; (b) two cascades of jointly and separately trained reference image compression model and GRDN~\cite{Kim_GRDN_2019}.}
\label{fig:comparison_QE}
\end{figure}
Fig.~\ref{fig:joint_opt_framework} shows our end-to-end joint learning scheme, JointIQ-Net, of image compression and quality enhancement in cascade. As mentioned, in this paper, we propose a novel image compression network but adopt an existing image quality enhancement network for the JointIQ-Net. Consequently, the proposed architecture provides high flexibility and extensibility. In particular, our method can easily accommodate future's advanced image quality enhancement networks, and it also allows various combinations of image compression and quality enhancement methods. That is, separately developed image compression networks and quality enhancement networks can easily be combined and can jointly be optimized in a unified architecture by minimizing the following total loss:
\begin{align}
\label{eq:total_loss_IQ}
\mathcal{L} = R + \lambda D(x, Q(I(\bm x)))
\end{align}
where $I$ is an image compression with input $\bm x$, and $Q$ is a quality enhancement function with input $\bm {\hat {x}}=I(\bm x)$ which is an intermediate reconstruction output of $I$. $R$, $D$, and $\lambda$ represent the rate, distortion, and a balancing parameter, respectively. In contrast to the previous methods~\cite{Balle17,Theis17,Balle18,Minnen2018,Lee2019} that train the image compression networks, $I$, to reconstruct the output images with as small distortion as possible, we regard the outputs of $I$ in Eq. (\ref{eq:total_loss_IQ}) as an intermediate latent representation, $\bm{\hat x}$, which is fed into the quality enhancement sub-network $Q$. So, the distortion $D$ is measured between the input $\bm x$ and the final output $\bm x' = Q(\bm {\hat {x}})$ reconstructed by $Q$. Consequently, our architecture allows two sub-networks to be jointly optimized towards minimizing the total loss Eq. (\ref{eq:total_loss_IQ}). Note that $\bm {\hat {x}}$ is best represented in a sense that $Q$ outputs the final reconstruction with high fidelity.

It should be noted that our work is not intended to propose an customized quality enhancement network, but to present an joint end-to-end learning scheme of both image compression and quality enhancement. Thus, to choose an appropriate quality enhancement network for our experiments, we combine a reference image compression method~\cite{Lee2019} with various quality improvement methods, VDSR~\cite{Kim_2016_VDSR}, RDN~\cite{Zhang_2018_CVPR} and GRDN~\cite{Kim_GRDN_2019}, in cascade connections. For fair comparisons, the numbers of parameters and layers for the quality enhancement networks are adjusted to have similar computation complexities. Fig. ~\ref{fig:comparison_QE} show the coding efficiency results for the combined image compression and quality enhancement networks. The experimental results are obtained by measuring average PSNR or MS-SSIM values over the Kodak PhotoCD image dataset \cite{KODAK}. In the supplementary material, we also provide the test results over the CLIC~\cite{CLIC} validation imageset and Tecnick~\cite{Asuni} imagesets.
 As shown in Fig. ~\ref{fig:comparison_QE}, the
GRDN~\cite{Kim_GRDN_2019} yields the highest compression performance in combination with the image compression method. So, we use GRDN~\cite{Kim_GRDN_2019} for our JointIQ-Net.

To verify the effectiveness of our joint learning scheme of image compression and quality enhancement, we compare two cascaded versions of the reference image compression model~\cite{Lee2019} and GRDN~\cite{Kim_GRDN_2019}. The first cascaded version is optimized for image compression and quality enhancement in an end-to-end manner, whereas the reference image compression model and the GRDN are separately learned in the second cascaded version where the GRDN is trained with the outputs of a separately trained reference image compression network for the same training dataset. 
Fig. ~\ref{fig:comparison_QE}-(b) shows the PSNR performance for the two cascaded versions. As shown, the first cascaded version outperforms the seconded version, especially in higher bit-rate ranges. The jointly trained GRDN in the first cascaded version effectively works over the whole bit-rate range while the separately trained GRDN can restore the visual quality of reconstructed images in low bit-rate ranges but improves less in high bit-rate ranges. 

It might be viewed that the quality enhancement network can be thought of as a decoder part of the image compression network. It can also be thought that increasing the decoder complexity of the image compression network might bring  a similar amount of coding efficiency improvement instead of cascading the quality enhancement network. However, the purpose of our work is targeted for a flexible joint learning scheme of image compression and any image quality enhancement solution that can be independently developed outside the image compression. Simply increasing the decoder part complexity may not bring a meaningful coding efficiency improvement due to its limited structure.  



\subsection{Proposed image compression network}
\begin{figure*}[t]
\begin{center}
\includegraphics[width=0.9\linewidth]{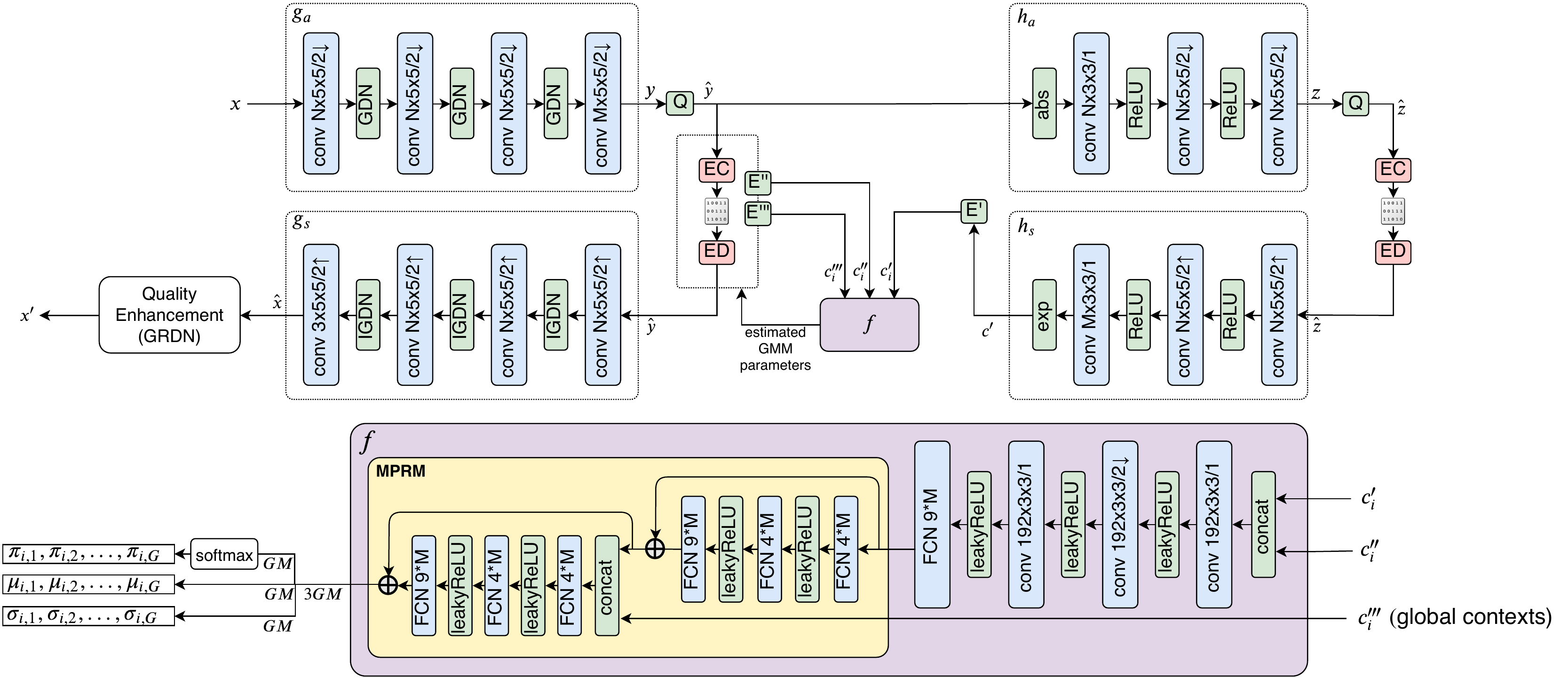}
\end{center}
   \caption{Architecture of our JointIQ-Net. Each convolutional layer is represented as the number of filters $\times$ filter height $\times$ filter width / the downscale or upscale factor, where $\uparrow$ and $\downarrow$ denote the up and down scaling via transposed convolutions, respectively. Input images are normalized into a scale between -1 and 1. $N$ and $M$ in the convolution layers indicate the numbers of feature map channels, while $M$ in each fully-connected layer is the number of nodes multiplied by its accompanying integer.}
\label{fig:overall_architecture}
\end{figure*}

The overall network architecture of JointIQ-Net is illustrated in Fig.~\ref{fig:overall_architecture}. As mentioned in section \ref{sec:Joint_opt}, our image compression network is connected with GRDN, adopted as a quality enhancement sub-network, in a cascade. The image compression network of the proposed JointIQ-Net is based on the existing approach~\cite{Lee2019}. Therefore, we basically use the same rate-distortion optimization framework and transform functions. The JointIQ-Net transforms input $\bm x$ into latent representations $\bm y$, and $\bm y$ is then quantized into $\bm{\hat y}$. In addition, we also use the hyperprior $\bm{\hat z}$, proposed in Ball{\'{e}} \etal(2018)~\cite{Balle18}’s approach, which further captures spatial correlations of $\bm{\hat y}$. Accordingly, we use four fundamental transform functions: an analysis transform $g_a(\bm x; \bm \phi_g)$, a synthesis transform $g_s(\bm{\hat y}; \bm \theta_g)$, an analysis transform $h_a(\bm {\hat y}; \bm \phi_h)$, and a synthesis transform $h_s(\bm {\hat z}; \bm \theta_h)$, as in the previous methods~\cite{Lee2019,Balle18}. The optimization process ensures the JointIQ-Net to yield the entropy of $\bm{\hat y}$ and $\bm{\hat z}$ as low as possible and also to yield $x'$, reconstructed from $\bm{\hat y}$, as close to the original visual quality as possible. To allow this rate-distortion optimization, along with the distortion between the input $\bm x$ and output $\bm x'$, the rate is calculated based on the prior probability models for $\bm{\hat y}$ and $\bm{\hat z}$. For $\bm{\hat z}$, we use a simple zero-mean Gaussian model convolved with $\mathcal U\bigl(-\tfrac 1 2, \tfrac 1 2\bigr)$, whose standard deviation values are found from training, whereas the parameters of the prior probability model for $\bm{\hat y}$ are estimated by the model parameter estimator $f$ in an autoregressive manner as in the previous method~\cite{Lee2019}. 

The model parameter estimator $f$ in the previous method~\cite{Lee2019} utilizes the two types of contexts, $\bm {c'}_{\bm i}$ reconstructed from the hyperprior $\bm{\hat z}$ and $\bm {c''}_{\bm i}$ extracted from the adjacent known representations of $\bm{\hat y}$. In addition, we let $f$ additionally utilize a global context, denoted as $\bm {c'''}_{\bm i}$, for estimating the model parameters more precisely as described in Section \ref{sec:global_context}. The functions to extract the three types of contexts are denoted as $E'$, $E''$, and $E'''$, respectively. With the three given contexts, $f$ estimates the parameters of GMM (convolved with $\mathcal U\bigl(-\tfrac 1 2, \tfrac 1 2\bigr)$), adopted as a prior probability model for $\bm{\hat{y}}$ in our method, as described in Section \ref{sec:Entropy_model}. This parameter estimation is used in the entropy coding and decoding processes, represented as EC and ED, as well as in the calculation of the rate term for training. In addition, we enhance the structure of the model estimator $f$, based on Lee \etal~\cite{Lee2019}'s method, by extending it to a new model estimator. The new model estimator incorporates a model parameter refinement module (MPRM) to improve the capability of model parameter estimation, as shown in Fig.~\ref{fig:overall_architecture}. The MPRM has two residual blocks, each of which contains the fully-connected layers and the corresponding non-linear activation layers.

\section{Improved entropy models and parameter estimation for entropy-minimization}
The previous entropy-minimization methods~\cite{Lee2019,Minnen2018} utilize local contexts to estimate the prior model parameters for each $\hat{y}_{\bm i}$. For this, they utilize the neighbor latent representations of a current representation, $\hat{y}_{\bm i}$, for estimating $\mu_{\bm i}$ and $\sigma_{\bm i}$ of a single Gaussian prior model (convolved with a uniform function) for $\hat{y}_{\bm i}$. These approaches have two limitations: (i) A single Gaussian model has a limited capability to model various distributions of latent representations. In this paper, we use a Gaussian mixture model GMM; (ii) Extracting the context information from neighbor latent representations is limited when their correlations widespreadly exist over the entire spatial domains.

\subsection{Gaussian mixture model (GMM) for prior distributions}
\label{sec:Entropy_model}
The existing autoregressive methods \cite{Lee2019,Minnen2018} use the single Gaussian distribution to model the distribution of each ${\hat y}_{\bm i}$. Although their transform networks can produce the latent representations that follow single Gaussian distributions, such a single Gaussian modeling is limited in predicting the actual distributions of latent representations, thus leading to sub-optimal performance. Instead, we use a more generalized form, GMM, of a prior probability model. 

\subsection{Formulation for Entropy Models}
We basically use the same R-D optimization framework as the existing approaches~\cite{Lee2019,Minnen2018}. The objective function includes the rate and distortion terms, as shown in Eq. \ref{eq:loss}, and the parameter $\lambda$ is used to adjust the balance between the rate and distortion in the optimization process:
\begin{align}
\label{eq:loss}
~&~~~~~~~~~~~~~~~~~~~\mathcal{L} = R + \lambda D \\
\text{with  } R &= \E_{\bm x \sim p_{\bm x}} \E_{\bm{\tilde y}, \bm{\tilde z} \sim q} \Bigl[- \log p_{\bm{\tilde y} \mid \bm{\hat z}}(\bm{\tilde y} \mid \bm{\hat z}) - \log p_{\bm{\tilde z}}(\bm{\tilde z})\Bigr], \notag \\
D &= \E_{\bm x \sim p_{\bm x}} \Bigl[- \log p_{\bm x \mid \bm{\hat y}}(\bm x \mid \bm{\hat y}) \Bigr] \notag
\end{align}
The rate term is composed of the cross-entropy for $\bm{\tilde z}$ and $\bm{\tilde y}|\bm{\hat z}$. To deal with the discontinuities due to quantization, as in the previous methods \cite{Balle18,Lee2019,Minnen2018}, a density function convolved with a uniform function $\mathcal U\bigl(-\tfrac 1 2, \tfrac 1 2\bigr)$ is used for approximating probability mass function (PMF) of $\bm{\hat y}$. Correspondingly, for training, the noisy representations $\bm{\tilde y}$ and $\bm{\tilde z}$ following uniform distributions whose mean values are $\bm{y}$ and $\bm{z}$, respectively, are used to fit the actual sample distributions to the PMF-approximating functions. To model the distributions of $\bm{\tilde z}$, as in previous approach~\cite{Lee2019}, we simply use zero-mean Gaussian density functions (convolved with a uniform density function), whose standard deviations are optimized via training.
Whereas, we extend the entropy model for $\bm{\tilde y}|\bm{\hat z}$ based on a GMM as:
\begin{align}
\label{eq:yhat_entropymodel}
p_{\bm{\tilde y} \mid \bm{\hat z}}(\bm{\tilde y} {\mid} \bm{\hat z}) {=} {\prod_{\bm i}} {\Bigl(} \sum_{g=1}^{G}\pi_{\bm{i},g}\mathcal N\bigl(\mu_{\bm{i},g}, \sigma_{\bm{i},g}^2\bigr) \ast \mathcal U\bigl(\text{-} \tfrac 1 2, \tfrac 1 2\bigr) {\Bigr)}(\tilde y_{\bm i})\\
\text{with  } \big\{\pi_{\bm{i},g}, \mu_{\bm{i},g}, \sigma_{\bm{i},g}|1 \leq g \leq G\big\} = f(\bm {c'}_{\bm i}, \bm {c''}_{\bm i}, \bm {c'''}_{\bm i})\notag
\end{align}
where $G$ is the number of Gaussian distribution functions. The distribution estimator $f$ predicts $3 \times G$ parameters so that each of the $G$ Gaussian distributions has its own weight, mean, and standard deviation parameters, denoted as $\pi_{\bm i,g}$, $\mu_{\bm i,g}$, and $\sigma_{\bm i,g}$, respectively.
The mean squared error (MSE) is basically used as the distortion term for optimization in Eq. \ref{eq:loss}, and we additionally provide experimental results of the MS-SSIM \cite{Wang03} optimized models.

\subsection{Global context for model parameter estimation}
\label{sec:global_context}
\begin{figure}[t]
\begin{center}
\includegraphics[width=0.7\linewidth]{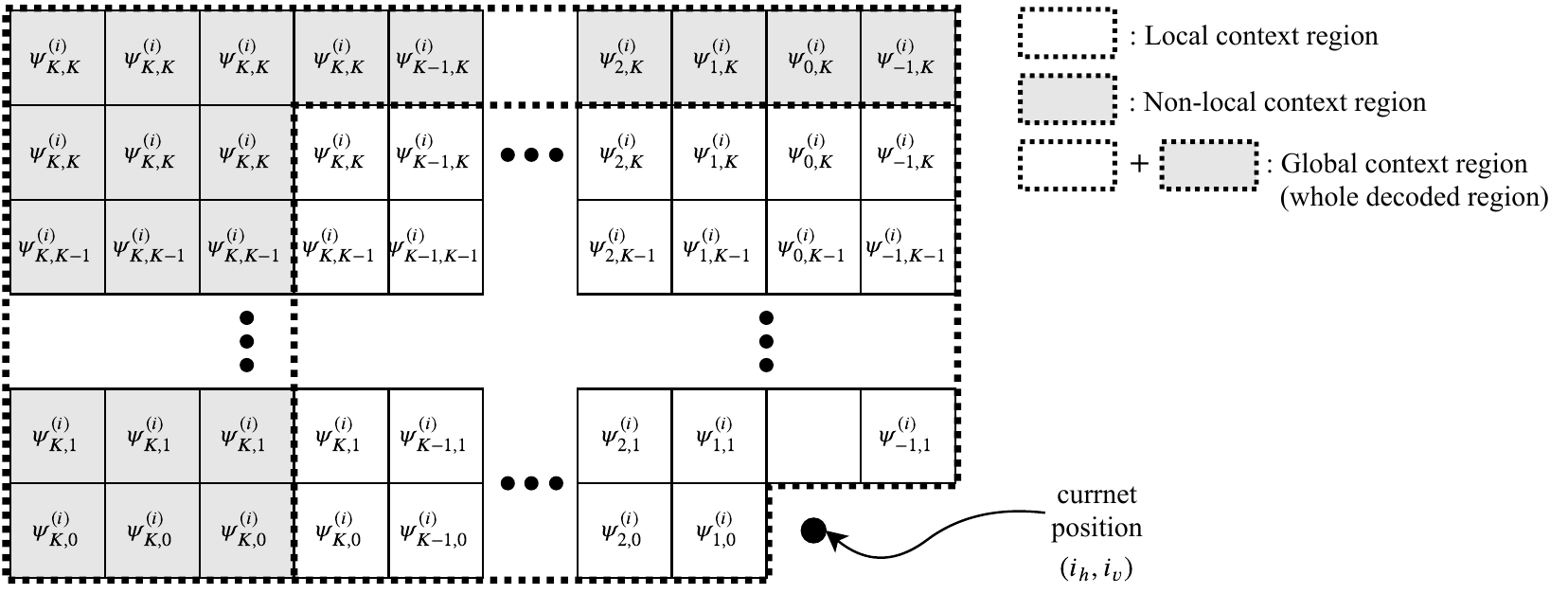}
\end{center}
   \caption{An example $\bm {a^{\bm{(i)}}}$, a set of ${\bm \psi^{\bm{(i)}}}$ variables mapped to the global context region. The softmax operation is then applied to $\bm {a^{\bm{(i)}}}$ to obtain the normalized weight $\bm{w}^{\bm{(i)}}$}
\label{fig:global_context_region}
\begin{center}
\includegraphics[width=1.0\linewidth]{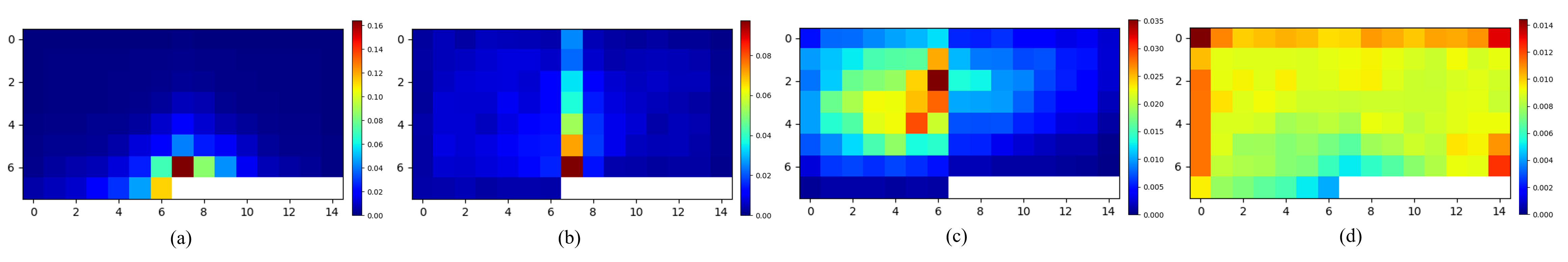}
\end{center}
   \caption{Examples of the trained $\bm{\psi}^{(\bm{i})}$, each of which is s set of weights for spatially aggregating contextual information from the whole spatial area of a specific channel of $\bm {\dot y}$. When a particular spatial position of $\bm {\dot y}$ is not covered by $\bm{\psi}^{(\bm{i})}$ because of $\bm{\psi}^{(\bm{i})}$'s limited size, the nearest weight value in $\bm{\psi}^{(\bm{i})}$ is shared for that spatial position. $\bm {\dot y}$ is a linearly transformed version of $\bm{\hat{y}}$ via an 1$\times$1 convolution layer.}
\label{fig:psi_visulization}
\end{figure}
In order to better extract context information for the current latent representation $y_{\bm i}$, we can use a global context by aggregating all possible contexts from the whole area of known representations to estimate the prior model parameters. For this, we define the global context as information aggregated from both local and non-local context regions, where the local context region is within the fixed distance, denoted as $K$, from the current representation $y_{\bm i}$, and the non-local region is the whole causal area outside the local context region. Fig. 6 shows an example of the local and non-local context regions.

As global context $\bm{c'''}_{\bm{i}}$, we use a weighted mean value and a weighted standard deviation value aggregated from the global context region, which is the whole known spatial area within a channel of $\bm {\dot y}$. We obtain the global context $\bm{c'''}_{\bm{i}}$ from $\bm {\dot y}$, which is a linearly transformed version of $\bm{\hat{y}}$ via an 1$\times$1 convolutional layer, rather than directly from $\bm {\hat y}$, to capture the correlations across the different channels of $\bm {\hat y}$ as well.
Specifically, The global context $\bm{c'''}_{\bm{i}} = \big\{\mu^{*}_{\bm{i}},\sigma^{*}_{\bm{i}}\big\}$ consists of a weighted mean $\mu^{*}_{\bm i}$ and a weighted standard deviation $\sigma^{*}_{\bm i}$, both of which are defined as: 

\begin{align}
\label{eq:global_context_mean}
&\mu^{*}_{\bm i}=\sum_{k,l\in{S}}w_{k,l}^{(\bm{i})}\dot{y}_{i_h\text{-}k,i_v\text{-}l}^{(\bm{i})}
\end{align}
\begin{align}
\label{eq:global_context_std}
&{\sigma^{*}_{\bm i}}= \sqrt{\frac{\sum_{k,l\in{S}}w_{k,l}^{(\bm{i})}{(\dot{y}_{i_h\text{-}k,i_v\text{-}l}^{(\bm{i})}-\mu^{*}_{\bm i})}^2}{1-\sum_{k,l\in{S}}{w_{k,l}^{(\bm{i})}}^2}}
\end{align}
where $\bm{i} = [{i_c,i_h,i_v}]$ is a 3-$d$ spatio-channel-wise position index indicating a current position $(i_h,i_v)$ in the $i_c$-$\textit{th}$ channel. $w_{k,l}^{(\bm{i})}$ is a weight variable for the relative coordinates $(k,l)$ based on the current location $(i_h,i_v)$, and $\dot{y}_{i_h\text{-}k,i_v\text{-}l}^{(\bm{i})}$ is a representation of $\bm{\dot{y}}^{(\bm i)}$ at location $(i_h\text{-}k,i_v\text{-}l)$, within the global context region $S$. $\bm{\dot{y}}^{(\bm i)}$ is the two-dimensional representaions within $i_c$-$\textit{th}$ channel of $\bm {\dot{y}}$. The weight variables in $\bm{w}^{(\bm i)}$ are the normalized weights that are element-wise multiplied to $\bm{\dot{y}}^{(\bm i)}$ for the weighted mean in Eq. \ref{eq:global_context_mean} and to the difference squares of ${(\dot{y}_{i_h-k,i_v-l}^{(\bm{i})}-\mu^{*}_{\bm i})}$ in Eq. \ref{eq:global_context_std}. 

Here, the key issue is to find an optimal set of the weight variables $\bm{w}^{(\bm i)}$ at every location $\bm{i}$. To obtain $\bm{w}^{(\bm i)}$ from a fixed number of trainable variables $\bm{\psi}^{(\bm i)}$, $\bm{w}^{(\bm i)}$ is estimated based on a 2-dimensional extension to the 1-dimensional global context extraction scheme of Shaw \etal~\cite{Shaw2018}’s approach. Fig. \ref{fig:global_context_region} shows the global context region that consists of the local context region within a fixed distance $K$, which is covered by the trainable variables $\bm{\psi}^{(\bm i)}$, and the non-local context region of a variable size, outside of the local context region. In the global context extraction, the non-local context region becomes enlarged as the local context window that defines the local context area slides over a feature map, thus increasing the number of weights $\bm{w}^{(\bm i)}$. To deal with the non-local context region which cannot covered by a fixed size of trainable variables $\bm{\psi}^{(\bm i)}$, a variable of $\bm{\psi}^{(\bm i)}$ allocated to the nearest local context area is used for each spatial position within the non-local context region, as shown in Fig. \ref{fig:global_context_region}. As a result, we can obtain a set of trainable variables $\bm{\psi}^{(\bm i)}$, denoted as $\bm {a^{\bm{(i)}}}$, corresponding to the global context region.
Then $\bm{w}^{(\bm i)}$ is calculated by normalizing $\bm {a^{\bm{(i)}}}$, via softmax as follows:
\begin{align}
\label{eq:weights}
\bm{w}^{\bm{(i)}} = softmax(\bm {a^{\bm{(i)}}})
\end{align}
where $\bm {a^{\bm{(i)}}}=\big\{\psi_{clip(k,K),clip(l,K)}^{(\bm{i})}|k,l \in{S} \big\}$ and $clip(x,$ $K)=max(-K,min(K,x))$. Note that $\psi_{k,l}^{(\bm{i})} = \psi_{k,l}^{(\bm{i}+\bm{c})}$ within the same channel (over the same spatial feature space). Fig.~\ref{fig:psi_visulization} visualizes the trained $\bm{\psi}^{(\bm{i})}$ examples for several channels of $\bm{\dot{y}}$. Fig.~\ref{fig:psi_visulization}-(a) shows the case that the context of the channel is dependent of the neighbor representations just next to the current latent representation while Fig.~\ref{fig:psi_visulization}-(d) shows the case that the context of the channel is dependent of the widely spread neighbor representations.

\section{Implementation}
The detailed structure of our JointIQ-Net is depicted in Fig.~\ref{fig:overall_architecture} where $N$ and $M$ are set according to $\lambda$ values. The values of $N$ and $M$ for different $\lambda$ values are tabulated in Table~\ref{table:hyperparameters}. We set $G$, the number of Gaussian pdfs for each prior distribution, to 3. Therefore, the model parameter estimator $f$ outputs $9 \times M$ values for $M$ representations 
of $\hat y_{\bm{i}}$, which are located at a specific spatial position of $\hat{\bm{y}}$. For obtaining the global contexts, we set $K$ to 7, and we utilize the global contexts only when the number of representations in the global context region is 30 or higher, in order to maintain statistical significance of the global contexts. For less than 30 representations, we set the global contexts to all zeros. For the GRDN in our final model, we set the number of GRDBs, RDBs in each GRDB, the number of convolutional layers in each RDB, and the number of kernels used by each convolutional layer to 4, 4, 8, and 64, respectively. Note that for the GDRN used in Fig.~\ref{fig:comparison_QE} and ~\ref{fig:ablation_study_results} is a light-weight version for which the above parameters are set to 4, 3, 3, and 32, respectively, for simulation at low complexity.

In the training phase, we used 51,140 256$\times$256 patches extracted from CLIC \cite{CLIC}  train imageset. The mini-batch size is set to 8, and all the models are trained using the ADAM optimizer~\cite{ADAM} with its default setting. The models were trained using their own initial learning rates in Table~\ref{table:hyperparameters}. We applied the gradient decaying by reducing the learning rate to half at every 50,000 steps during
the final 300,000 steps. For proper scaling of the $\lambda$ range, Eq.~\ref{eq:actual_loss} is used as an objective cost function in the actual implementation.

For training the final models in Fig.~\ref{fig:experimental_results}, which include the deeper $Q$ (GRDN) sub-networks, we used the same 256$\times$256-sized patches as a training set, but we randomly extracted 96$\times$96-sized patches from the outputs of the $I$ sub-network, and fed them into the $Q$ sub-network. The distortion term was calculated against the corresponding area of input patches, and the rate term was also calculated over the corresponding 6$\times$6-sized area out of the 16$\times$16-sized spatial area of $\hat{\bm {y}}$. To reduce the training time, we utilized the pre-trained $I$ sub-networks of the models with the lightweight GRDNs. We first train the $Q$ sub-network using only the distortion term for 100K iterations, and then further optimize the $I$ and $Q$ sub-networks in an end-to-end manner for additional 1M iterations.

\begin{eqnarray}
\label{eq:actual_loss}
	\mathcal{L} = \frac{\lambda}{W_y \cdot H_y \cdot 256} R
	+ \frac{1 - \lambda}{1000} D.
\end{eqnarray}

\setlength{\tabcolsep}{4pt}
\begin{table}
\begin{center}
\caption{Hyperparameters for the models trained with different $\lambda$ values.}
\label{table:hyperparameters}
\begin{tabular}{ccccc}
\hline\noalign{\smallskip}
$\lambda$ & $N$ & $M$ & No. of iterations & Initial learning rates\\
\noalign{\smallskip}
\hline
\noalign{\smallskip}
0.5	& 128 & 128 & 1.2M & 1e-4 \\
0.35 & 128 & 128 & 1.2M & 1e-4 \\
0.23 & 128 & 192 & 1.5M & 1e-4 \\
0.12 & 192 & 256 & 1.5M & 1e-4 \\
0.06 & 192 & 420 & 2.0M & 5e-5 \\
0.03 & 192 & 420 & 2.0M & 5e-5 \\
0.017 & 256 & 600 & 3.0M & 5e-5 \\
0.01 & 256 & 600 & 3.0M & 3e-5 \\
\hline
\end{tabular}
\end{center}
\end{table}
\setlength{\tabcolsep}{1.4pt}

\section{Experiments}
\label{sec:experiments}
\subsection{Experimental environments}
To verify the performance of the proposed method, we measured the average bits per pixel (BPP) and quality of the reconstructed images over the Kodak PhotoCD image dataset \cite{KODAK}. The PSNR and MS-SSIM metrics are used to measure the quantitative qualities. For each quality metric, eight models were trained with different $\lambda$ values, and we evaluated them by comparing the resulting R-D curve with those of the existing ANN-based approaches, such as Lee \etal~\cite{Lee2019}, Ball{\'{e}} \etal(2018)~\cite{Balle18}, and Theis \etal~\cite{Theis17}, and the conventional codecs, such as VVC Intra (VTM 7.1~\cite{VTM}), BPG \cite{BPG}, and JPEG2000~\cite{Taubman2001}. Minnen \etal~\cite{Minnen2018}'s method is also one of the representative ANN-based compression approaches, but was excluded for comparison because their method showed very similar performance to that of Lee \etal~\cite{Lee2019}'s approach. We compared the results in the range from 0.1 bpp to 1.5 bpp.

\subsection{Experimental results}
We compared the compression performance of our method with those of the other existing approaches using the rate-distortion curves, in terms of PSNR and MS-SSIM. As demonstrated in Fig.~\ref{fig:experimental_results}, our method outperforms all the previous methods under comparison in terms of both PSNR and MS-SSIM. Specifically, the compression gains are obtained with 1.65 (48.40)\%, 16.96 (14.83)\%, 26.58 (26.65)\%, 22.57 (57.35)\%, and 45.48 (73.65)\% in the BD-rates of PSNR (MS-SSIM) over VVC Intra (VTM 7.1~\cite{VTM}), Lee \etal~\cite{Lee2019}'s method, Ball{\'{e}} \etal(2018)~\cite{Balle18}'s method, BPG \cite{BPG} and JPEG2000 \cite{Taubman2001}, respectively. In the supplementary material, we provide the examples of reconstructed images with those of the other methods.

\subsection{Ablation study}
\label{sec:ablation_study}
\begin{figure}[t]
\begin{center}
\includegraphics[width=1.0\linewidth]{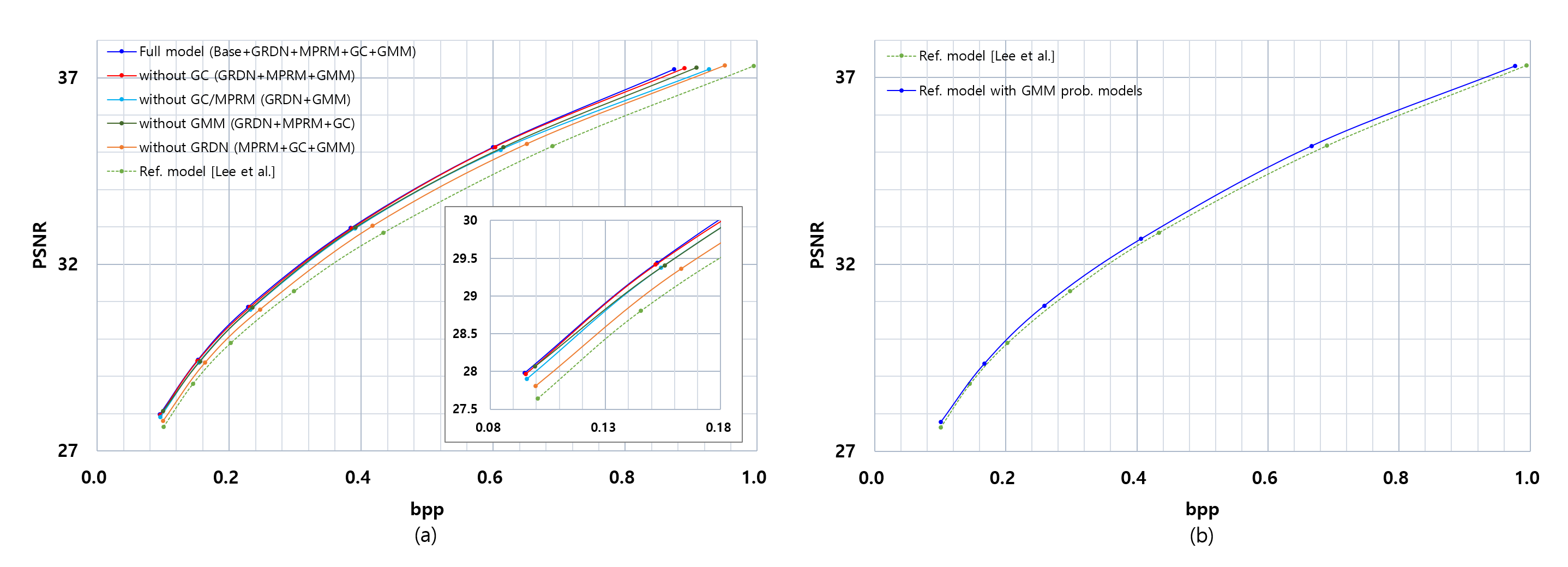}
\end{center}
   \caption{PSNR performance for Kodak PhotoCD \cite{KODAK}. (a) JointIQ-Net variations; (b) a reference~\cite{Lee2019} using single Gaussian prior models and the same model~\cite{Lee2019} with GMM prior models.}
\label{fig:ablation_study_results}
\end{figure}
In order to verify the effectiveness of each proposed element, we conducted the ablation study as follows: We excluded each proposed element from the full model, and trained the models in the same way as in the experiments of Section~\ref{sec:experiments}. We compared the test results of each model with those of the full model. Four different models were evaluated and their excluded components are GRDN~\cite{Kim_GRDN_2019}, global context (GC), model parameter refinement module (MPRM), and Gaussian mixture model (GMM), respectively. Note that, the global context is also excluded when excluding MPRM because the global context is processed by the MPRM in our full model. In addition, as a baseline, the compression performance of Lee \etal~\cite{Lee2019}’s method is also included in the comparison. Fig.~\ref{fig:ablation_study_results} (a) shows the PSNR performances for various versions of our model used in the ablation study. As shown in Fig.~\ref{fig:ablation_study_results} (a), when GRDN~\cite{Kim_GRDN_2019} is excluded, significant performance degradation occurs. This indicates that the proposed joint learning scheme can play an important role in improving compression performance. The global context also improves performance, but the amount of PSNR performance improvement is relatively low. When both MPRM and global context are excluded, the performance degradation becomes more noticeable. The results also show that MPRM has a greater impact in a higher bpp range. Whereas, when we use a single Gaussian model instead of a GMM, a similar level of performance degradation occurs over the entire bpp range. Table~\ref{table:ablation_study_results} compares the quantitative results between the final model and each of the element-excluded models in terms of BD-rate loss.

To see the effectiveness of a GMM prior model, we performed a comparison test between the reference model using a single Gaussian prior model and the same model with a GMM prior model, without the other architectural or contextual changes. As shown in Fig.~\ref{fig:ablation_study_results}, we've obtained 3.63\% of coding gain when using the GMM prior model compared to the reference model. Note that two networks have the same architecture, except for the number of the output nodes of the model estimator $f$.

\setlength{\tabcolsep}{4pt}
\begin{table}
\begin{center}
\caption{BD-rate losses for element-excluded models in comparison with the full model, JointIQ-Net (anchor).}
\label{table:ablation_study_results}
\begin{tabular}{cc}
\hline\noalign{\smallskip}
Models & BD-rate losses (\%)\\
\noalign{\smallskip}
\hline
\noalign{\smallskip}
without GRDN & 8.61 \\
without GC & 0.92 \\
without GC/MPRM & 3.71 \\
without GMM & 2.76 \\
\hline
\end{tabular}
\end{center}
\end{table}
\setlength{\tabcolsep}{1.4pt}

\section{Conclusion}
In this paper, we proposed a new image compression method, called JointIQ-Net, that outperforms VVC Intra (VTM 7.1), BPG, JPEG2000, and the state-of-the-art ANN-based image compression approaches. To the best of our knowledge, our JointIQ-Net is the \textit{first} learned image compression approach that surpasses the VVC Intra coding, in terms of both PSNR and MS-SSIM. For improving the coding efficiency of the JointIQ-Net, we have built a new joint learning scheme that incorporates both image compression and quality enhancement, allowing them to be end-to-end optimized in a unified manner. From the perspective of image compression, we have improved the entropy model by adopting GMM as a more generalized prior probability model for the transformed representations. In addition, we have enhanced the capability of the model parameter estimation by utilizing the global contexts that can reduce the remaining correlations over the global regions of the transformed representations.

\clearpage
%

%
%
\bibliographystyle{splncs04}
\bibliography{egbib}

\clearpage
\section{Supplementary material}
As denoted in the main paper, our JointIQ-Net is the \textit{first work} that surpasses the coding efficiency performance of the the most recent and advanced image compression coding, VVC Intra Coding (VTM 7.1~\cite{VTM}), that has been almost finalized for standardization. This Supplemental Material provides a plenty of experimental results that support the superiority of our JointIQ-Net against the state-of-the-art (SOTA) image compression methods. As shown, our JointIQ-Net has yielded \textit{a substantial improvement on coding efficiency} against the SOTA methods.   

\subsection{Experimental results on CLIC and Tecnick imagesets}
To thoroughly inspect the effectiveness of our JointIQ-Net, we further performed comparison experiments between our model and the SOTA methods including VVC Intra coding (VTM 7.1~\cite{VTM}), BPG~\cite{BPG}, and Lee \etal~\cite{Lee2019}'s approach, over two different image datasets, the CLIC~\cite{CLIC} validation set and the $\it{SAMPLING}$ test set (color format: RGB, bit depth: 8 bits per channel, resolution: 1200 $\times$ 1200) of Tecnick~\cite{Asuni} image set. 
\setlength{\tabcolsep}{4pt}
\begin{table}
\begin{center}
\caption{BD-rate gains of our JointIQ-Net against the VVC Intra coding (VTM 7.1~\cite{VTM}), BPG~\cite{BPG}, and Lee \etal~\cite{Lee2019}'s approach for CLIC~\cite{CLIC} validation set, and against the VTM 7.1~\cite{VTM}, BPG~\cite{BPG}, and Lee \etal~\cite{Lee2019}'s approach for Tecnick~\cite{Asuni} image set. The third row shows the coding gains of our MSE-optimized model in terms of PSNR versus BD-rate, and the fourth row indicates the coding gains of our MS-SSIM optimized model in terms of MS-SSIM versus BD-rate.}
\label{table:further_experimental_results}
\begin{tabular}{l ccc ccc}
\toprule
 & \multicolumn{3}{c}{CLIC~\cite{CLIC}} & \multicolumn{3}{c}{Tecnick~\cite{Asuni}} \\
\cmidrule(lr){2-4} \cmidrule(lr){5-7}
 & VVC Intra~\cite{VTM} & BPG~\cite{BPG} & Lee~\cite{Lee2019} & VVC Intra~\cite{VTM} & BPG~\cite{BPG} & Lee~\cite{Lee2019} \\
\midrule
MSE opt. & 4.85\% & 28.16\% & 21.03\% & 7.12\% & 35.93\% & 28.21\%\\
MS-SSIM opt. & 52.60\% & 62.19\% & 21.25\% & 37.85\% & 54.78\% & 21.97\%\\
\bottomrule
\end{tabular}
\end{center}
\end{table}
\setlength{\tabcolsep}{1.4pt}

Table~\ref{table:further_experimental_results} shows the BD-rate gains of our JointIQ-Net against the VTM 7.1~\cite{VTM}, BPG~\cite{BPG}, and Lee \etal~\cite{Lee2019}'s approach for CLIC~\cite{CLIC} validation set, and against the VTM 7.1~\cite{VTM}, BPG~\cite{BPG}, and Lee \etal~\cite{Lee2019}'s approach for Tecnick~\cite{Asuni} image set. It is clear in Table~\ref{table:further_experimental_results} that our JointIQ-Net significantly improves the coding efficiency over the SOTA methods, especially outperforming the the VTM 7.1~\cite{VTM} with average 4.85\% and 7.12\%, respectively, for the CLIC~\cite{CLIC} validation set and the Tecnick~\cite{Asuni} image set. It should be noted that we performed one line of padding to each input image (feature map) at a convolution layer when down-scaling is needed. That is, when the number of horizontal (vertical) lines in an input image (feature map) is odd and the input image (feature map) is needed to be down-scaled, one more horizontal (vertical) line is padded in the most bottom (the most right) to have an even line number before down-scaling. Correspondingly, the decoder removes the padded horizontal (vertical) line, at each scale, based on the transmitted original input size. Because the down-scaling and up-scaling architectures of the encoder and decoder are symmetric, the decoder can distinguish the unnecessary lines by emulating the padding area decision process of the encoder. In our experimental results, furthermore, the sizes of file headers indicating the original input sizes were included in bpp calculation.

\begin{figure}
\begin{center}
\includegraphics[width=1.0\linewidth]{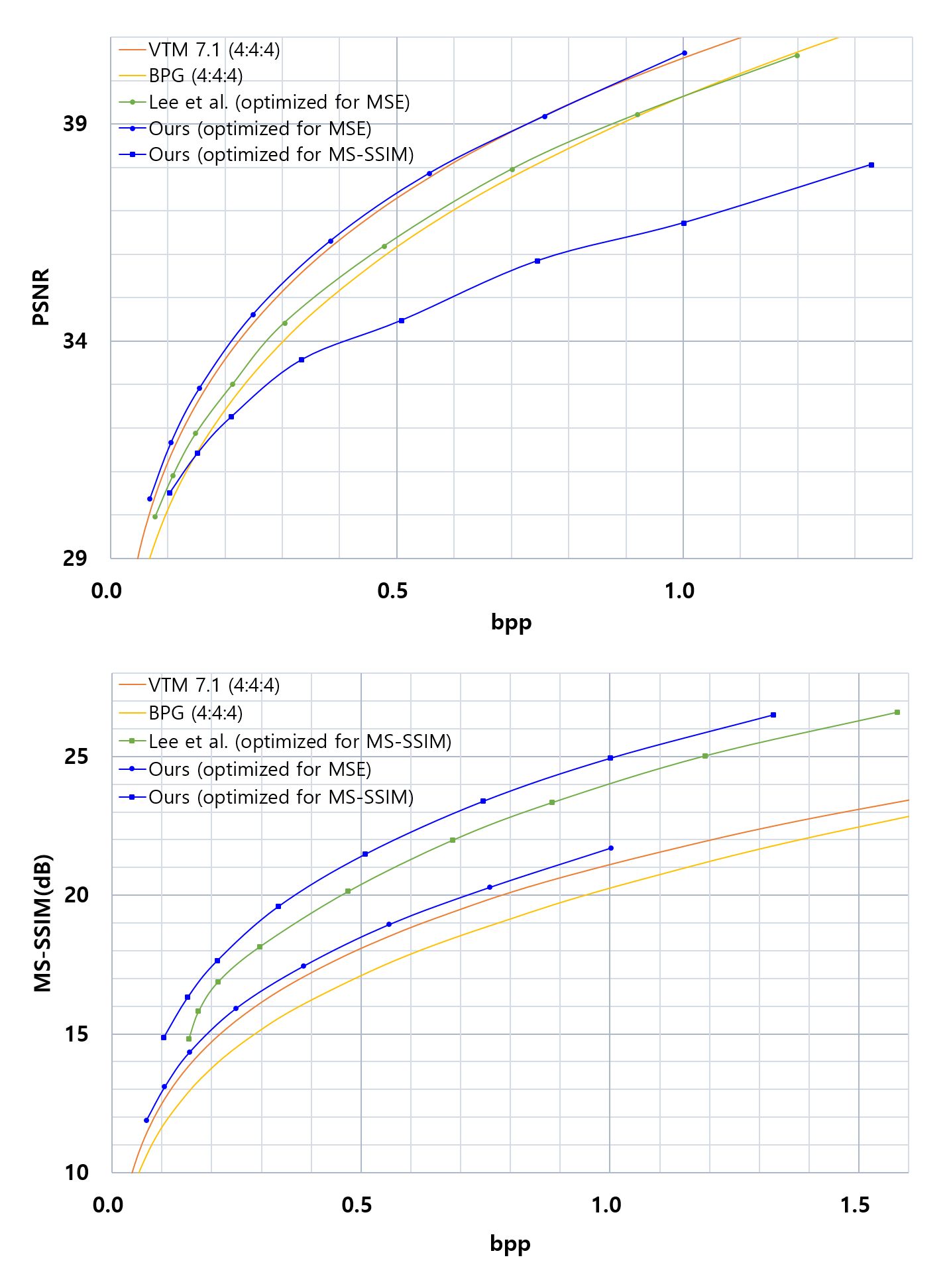}
\end{center}
   \caption{Rate-distortion curves of our JointIQ-Net and the SOTA methods, VTM 7.1~\cite{VTM}, BPG~\cite{BPG}, for the CLIC validation image dataset \cite{CLIC}. The top and bottom plots represent RD-curves in terms of PSNR and MS-SSIM, respectively.}
\label{fig:experimental_results_CLIC}
\end{figure}

\begin{figure}
\begin{center}
\includegraphics[width=1.0\linewidth]{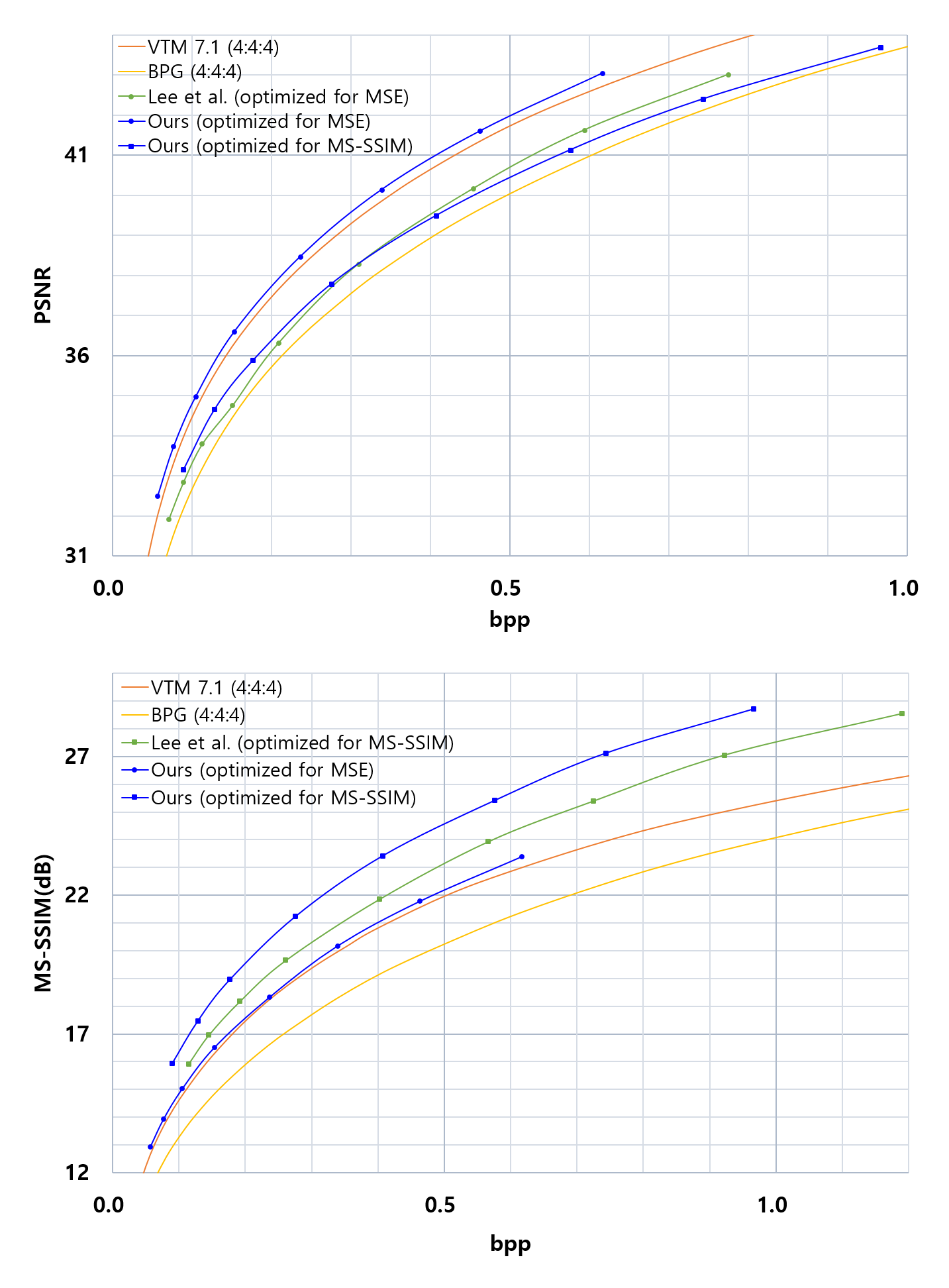}

\end{center}
   \caption{Rate-distortion curves of our JointIQ-Net and the SOTA methods, VTM 7.1~\cite{VTM}, BPG~\cite{BPG}, for the TECNICK image dataset \cite{Asuni}. The top and bottom plots represent RD-curves in terms of PSNR and MS-SSIM, respectively.} 
\label{fig:experimental_results_TECNICK}
\end{figure}

Fig.~\ref{fig:experimental_results_CLIC} and \ref{fig:experimental_results_TECNICK} show the coding efficiency curves for the results in Table~\ref{table:further_experimental_results} for the CLIC~\cite{CLIC} validation set and the $\it{SAMPLING}$ testset of Tecnick~\cite{Asuni} imageset, respectively. It should be noted in Fig.~\ref{fig:experimental_results_CLIC} and \ref{fig:experimental_results_TECNICK} that our JointIQ-Net outperforms all the SOTA methods over the entire bpp range.

\clearpage

\subsection{Subjective comparisons of the reconstructed images}
\label{sec:samples}%
\begin{figure*}[p]
	\vspace{-0.1cm}
	\hbox{
		\hspace{-0.1cm}
		\begin{tikzpicture}
			\node at (0.0cm, 8.9cm) {\includegraphics[width=5.8cm]{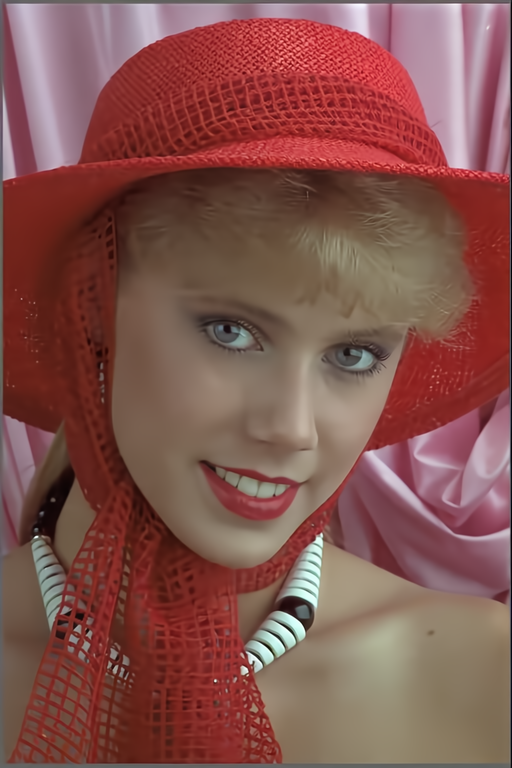}};
			\node at (6.0cm, 8.9cm) {\includegraphics[width=5.8cm]{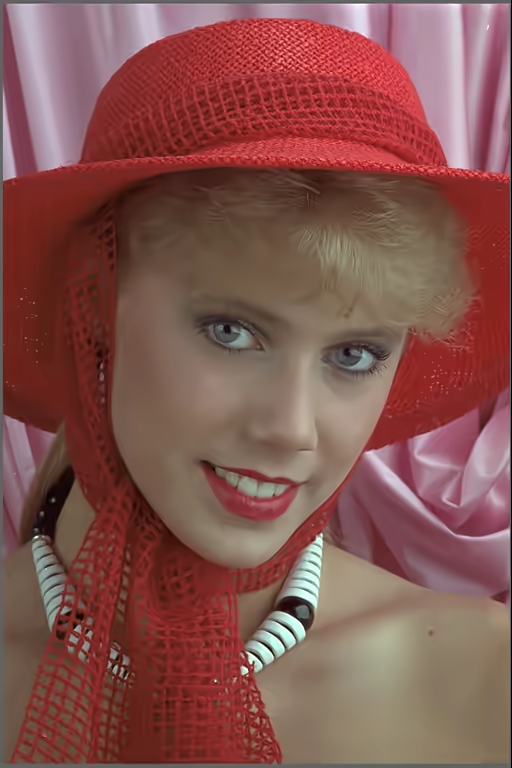}};
			\node at (0.0cm, 0.0cm) {\includegraphics[width=5.8cm]{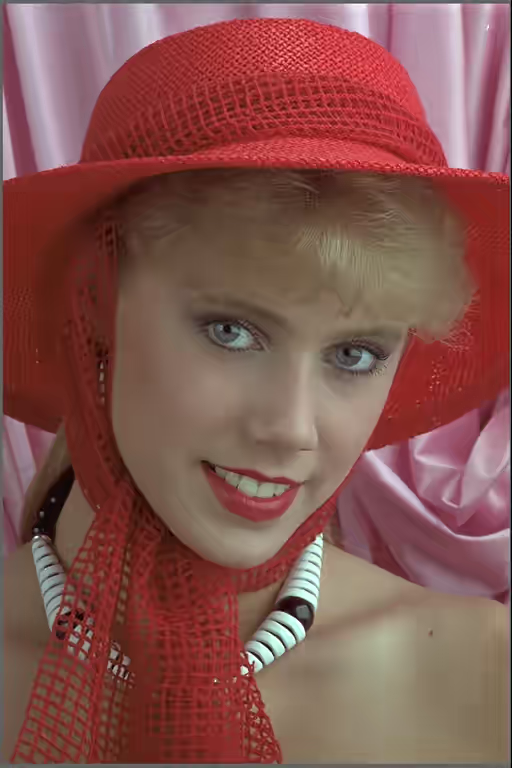}};
			\node at (6.0cm, 0.0cm) {\includegraphics[width=5.8cm]{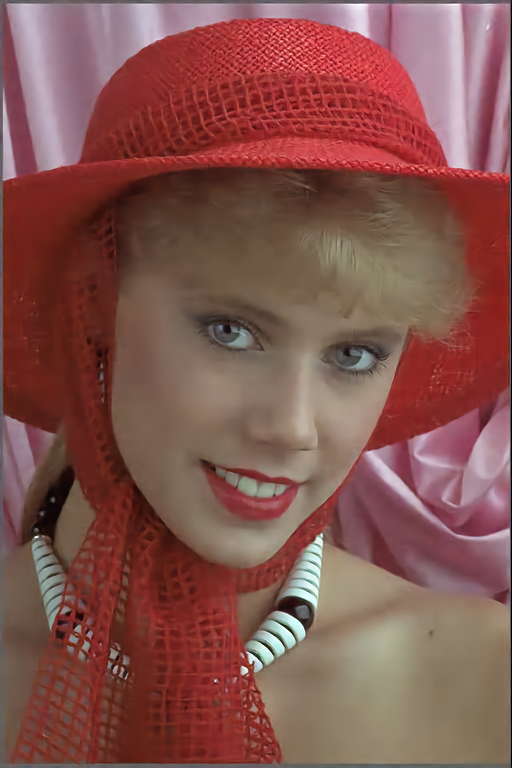}};
		\end{tikzpicture}
	}
	\captionsetup{width=0.97\linewidth}
	\caption{Subjective comparison of decoded images by our JointIQ-Net, VTM 7.1~\cite{VTM}, Lee \etal~\cite{Lee2019}'s approach, and BPG~\cite{BPG} in the clockwise order. Top-left, our JointIQ-Net (MSE-optimized; bpp, 0.2035; PSNR, 33.1097); top-right, VTM 7.1~\cite{VTM} (bpp, 0.2041; PSNR, 32.9663); bottom-left, BPG~\cite{BPG} (bpp, 0.2078; PSNR, 32.0406); bottom-right, Lee \etal~\cite{Lee2019}'s method (MSE-optimized; bpp, 0.2040; PSNR, 32.2065)}
\label{fig:supp1}
\end{figure*}

\begin{figure*}[p]
	\vspace{-0.1cm}
	\hbox{
		\hspace{-0.1cm}
		\begin{tikzpicture}
			\node at (0.0cm, 8.9cm) {\includegraphics[width=5.8cm]{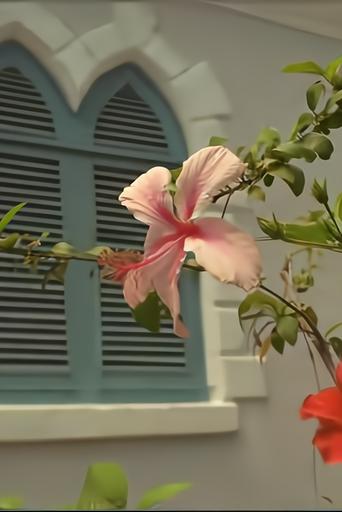}};
			\node at (6.0cm, 8.9cm) {\includegraphics[width=5.8cm]{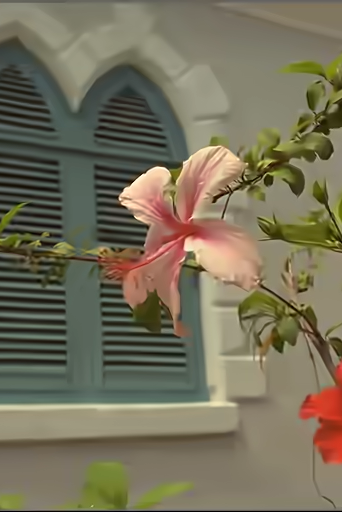}};
			\node at (0.0cm, 0.0cm) {\includegraphics[width=5.8cm]{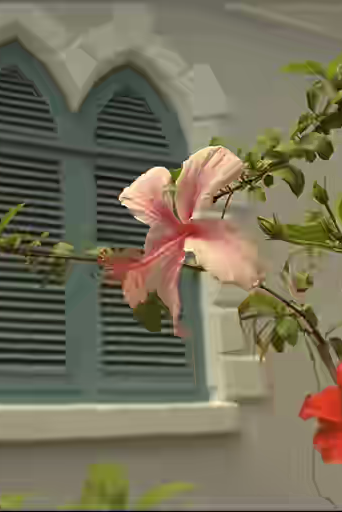}};
			\node at (6.0cm, 0.0cm) {\includegraphics[width=5.8cm]{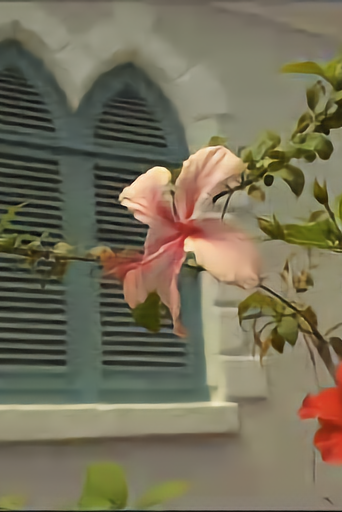}};
		\end{tikzpicture}
	}
	\captionsetup{width=0.97\linewidth}
	\caption{Subjective comparison of decoded images by our JointIQ-Net, VTM 7.1~\cite{VTM}, Lee \etal~\cite{Lee2019}'s approach, and BPG~\cite{BPG} in the clockwise order. Top-left, our JointIQ-Net (MSE optimized; bpp, 0.1243; PSNR, 31.3978); top-right, VTM 7.1~\cite{VTM} (bpp, 0.1248; PSNR, 30.9643); bottom-left, BPG~\cite{BPG} (bpp, 0.1188; PSNR, 29.4102); bottom-right, Lee \etal~\cite{Lee2019}'s method (MSE optimized; bpp, 0.1043; PSNR, 29.46)}
\label{fig:supp2}
\end{figure*}

Figs. \ref{fig:supp1} and \ref{fig:supp2} show the decoded images of KODIM04 and KODIM07 by our MSE-optimized JointIQ-Net, VTM 7.1~\cite{VTM}, Lee \etal~\cite{Lee2019}'s and BPG~\cite{BPG} in the clockwise order. As shown in Figs. \ref{fig:supp1} and \ref{fig:supp2}, the decoded images by our JointIQ-Net look visually more pleasing over the other decoded images by the SOTA methods.

\begin{figure*}[p]
	\vspace{-0.1cm}
	\hbox{
		\hspace{-0.1cm}
		\begin{tikzpicture}
			\node at (0.0cm, 8.9cm) {\includegraphics[width=5.8cm]{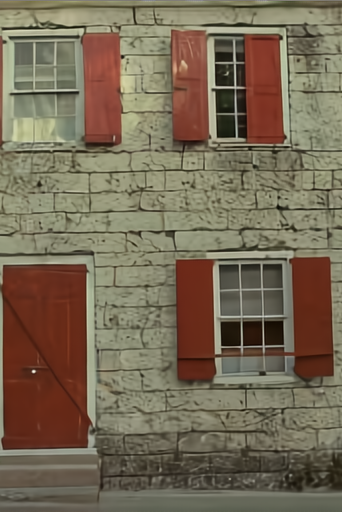}};
			\node at (6.0cm, 8.9cm) {\includegraphics[width=5.8cm]{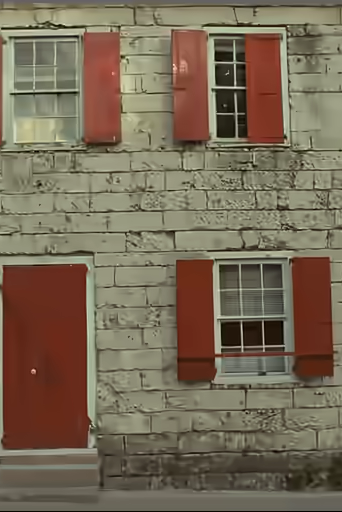}};
			\node at (0.0cm, 0.0cm) {\includegraphics[width=5.8cm]{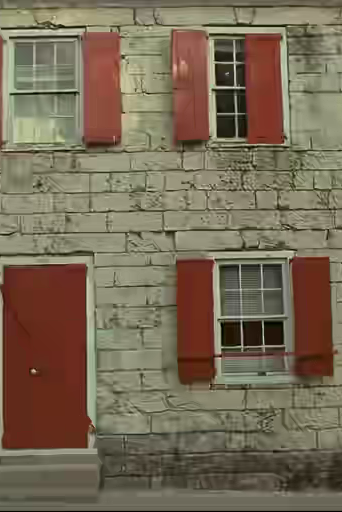}};
			\node at (6.0cm, 0.0cm) {\includegraphics[width=5.8cm]{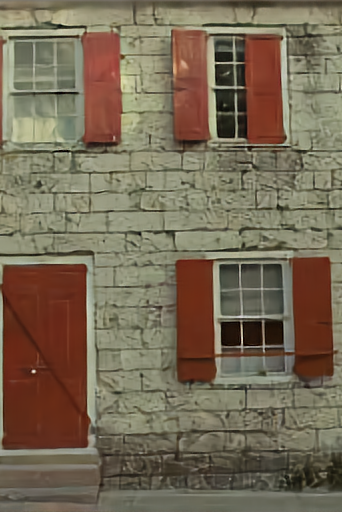}};
		\end{tikzpicture}
	}
	\captionsetup{width=0.97\linewidth}
	\caption{Subjective comparison of decoded images by our JointIQ-Net, VTM 7.1~\cite{VTM}, Lee \etal~\cite{Lee2019}'s approach, and BPG~\cite{BPG} in the clockwise order. Top-left, our JointIQ-Net (MS-SSIM optimized; bpp, 0.2004; MS-SSIM, 0.9528); top-right, VTM 7.1~\cite{VTM} (bpp, 0.1978; MS-SSIM, 0.9278); bottom-left, BPG~\cite{BPG} (bpp, 0.1920; MS-SSIM, 0.9136); bottom-right, Lee \etal~\cite{Lee2019}'s method (MS-SSIM optimized; bpp, 0.2073; MS-SSIM, 0.9488)}
\label{fig:supp5}
\end{figure*}

\begin{figure*}[p]
	\vspace{-0.1cm}
	\hbox{
		\hspace{-0.1cm}
		\begin{tikzpicture}
			\node at (0.0cm, 8.9cm) {\includegraphics[width=5.8cm]{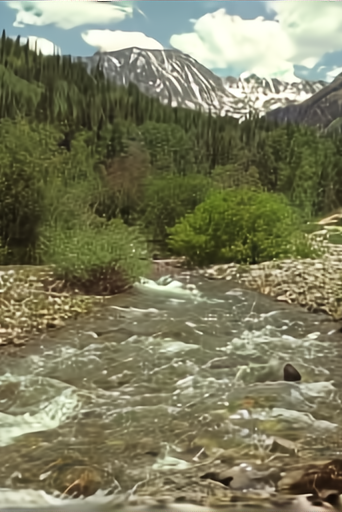}};
			\node at (6.0cm, 8.9cm) {\includegraphics[width=5.8cm]{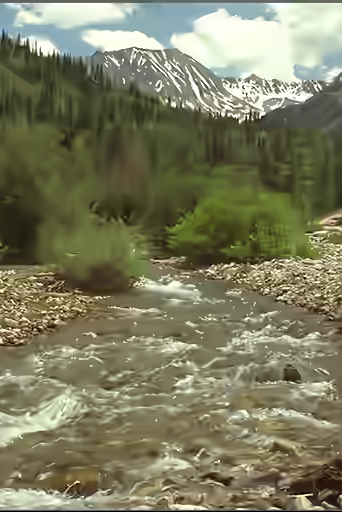}};
			\node at (0.0cm, 0.0cm) {\includegraphics[width=5.8cm]{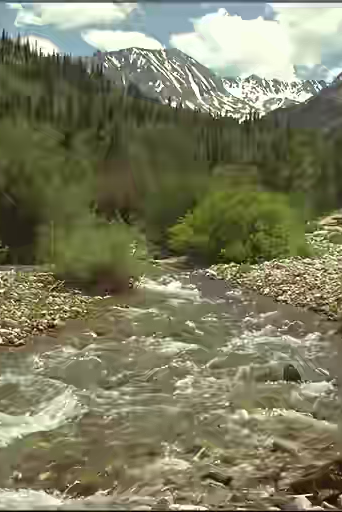}};
			\node at (6.0cm, 0.0cm) {\includegraphics[width=5.8cm]{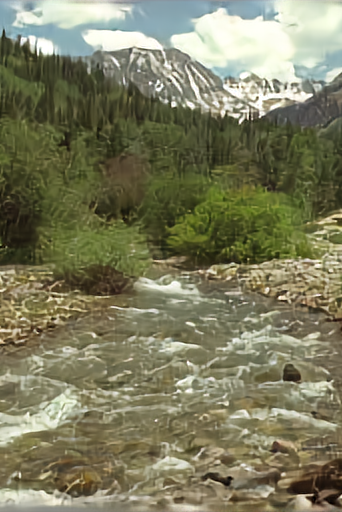}};
		\end{tikzpicture}
	}
	\captionsetup{width=0.97\linewidth}
	\caption{Subjective comparison of decoded images by our JointIQ-Net, VTM 7.1~\cite{VTM}, Lee \etal~\cite{Lee2019}'s approach, and BPG~\cite{BPG} in the clockwise order. Top-left, our JointIQ-Net (MS-SSIM optimized; bpp, 0.2442; MS-SSIM, 0.9319); top-right, VTM 7.1~\cite{VTM} (bpp, 0.2409; MS-SSIM, 0.8719); bottom-left, BPG~\cite{BPG} (bpp, 0.2760; MS-SSIM, 0.8699); bottom-right, Lee \etal~\cite{Lee2019}'s method (MS-SSIM optimized; bpp, 0.2630; MS-SSIM, 0.9313)}
\label{fig:supp7}
\end{figure*}

Figs. \ref{fig:supp5} and \ref{fig:supp7} are the decoded images of KODIM01 and KODIM13 by our MS-SSIM-optimized JointIQ-Net, VTM 7.1~\cite{VTM}, LEE \etal~\cite{Lee2019}'s and BPG~\cite{BPG} in the clockwise order. As also shown in \ref{fig:supp5} and \ref{fig:supp7}, our JointIQ-Net yields the decoded images with better perceptual quality over the other decoded images by the SOTA methods. Note that we cropped the both sides of the decoded images to fit the page width in the case of the horizontally long images (KODIM01, KODIM07, and KODIM13) for convenient visualization.

\end{document}